\documentclass[a4paper]{revtex4}
\usepackage{graphicx}
\usepackage{fancyhdr}
\usepackage{amsmath}
\usepackage{color}
\graphicspath{{figs/}}

\newcommand{\MHexp}{125}

\input{paperdef}

\begin{document}

\title{The MSSM Higgs Sector at the LHC and Beyond}

%

\author{S.\ Heinemeyer}
\affiliation{Instituto de F\'isica de Cantabria (CSIC), E--39005 Santander,
  Spain} 

\begin{abstract}
Some possibilities to test the Higgs sector of the Minimal Supersymmetric
Standard Model (MSSM) at the LHC and future $e^+e^-$ colliders are
discussed. This includes precision coupling strength measurements, the
search for additional Higgs bosons as well as their decay to
supersymmetric particles.
\end{abstract}

\maketitle

\thispagestyle{fancy}



\section{Introduction}

The ATLAS and CMS experiments at CERN have discovered a
new boson with a mass around 
$\MHexp \gev$~\cite{ATLASdiscovery,CMSdiscovery}. 
Within the present experimental uncertainties
this new boson behaves like the 
Higgs boson of the Standard Model (SM)~\cite{Moriond15}. However,
the newly discovered 
particle can also be interpreted as the Higgs boson of extended models,
where the  Minimal Supersymmetric
Standard Model (MSSM)~\cite{mssm} is a prime candidate. 
The Higgs sector of the MSSM with two scalar doublets
accommodates five physical Higgs bosons. In
lowest order these are the light and heavy $\cp$-even $h$
and $H$, the $\cp$-odd $A$, and the charged Higgs bosons $H^\pm$.
It can be expressed (at lowest order) in terms of the gauge couplings,
the mass of the $\cp$-odd Higgs boson, 
$\MA$, and $\tb \equiv v_2/v_1$,  
the ratio of the two vacuum expectation values. All other masses and
mixing angles can therefore be predicted.
Higher-order contributions can give 
large corrections to the tree-level relations~~\cite{MHreviews,PomssmRep}. 
An upper bound for the
mass of the lightest MSSM Higgs boson of $\Mh \lsim 135 \gev$  had been
obtained~\cite{mhiggsAEC}, in perfect agreement with the observed value
of 
\begin{align}
\MH^{\rm exp} &= 125.09 \pm 0.24 \gev~,
\label{MHexp}
\end{align}
as evaluated by the combination of ATLAS and CMS
measurements~\cite{ATLASCMS-MH}. 

We will review a few ways to test the MSSM Higgs sector at the LHC and
beyond. We first briefly discuss the precision prediction for the
lightest $\cp$-even Higgs boson mass in the MSSM. We review the status
and the prospects of the Higgs coupling strength analyses at the LHC and
the ILC. Finally we discuss where additional MSSM Higgs bosons could be
discovered and review some precision calculations for their (potential)
decay to supersymmetric (SUSY) particles.


\section{The lightest Higgs boson mass as a precision observable}

In the MSSM the mass of the light
$\cp$-even Higgs boson, $\Mh$, can directly be predicted from
the other parameters of the model. The accuracy of this prediction 
should at least match the one of the experimental result.
The measured Higgs-boson mass value, \refeq{MHexp}, 
has already reached the level of a precision
observable with an experimental accuracy of about $250 \mev$.
Consequently, it plays an important role in the context of testing the
MSSM Higgs sector.

The status of higher-order corrections to $\Mh$ is quite advanced, see 
\citeres{ERZ,mhiggsf1lABC}
for the calculations of the full one-loop level.
At the two-loop level~\cite{mhiggsletter,mhiggslong,mhiggslle,mhiggsFD2,bse,mhiggsEP0,mhiggsEP1,mhiggsEP1b,mhiggsEP2,mhiggsEP3,mhiggsEP3b,mhiggsEP4,mhiggsEP4b,mhiggsRG1,mhiggsRG1a,mhiggsFDalt2}
in particular the \order{\alt\als} and \order{\alt^2} contributions 
($\alt \equiv h_t^2 / (4 \pi)$, $h_t$ being the top-quark Yukawa coupling)
to the self-energies -- evaluated in the 
Feynman-diagrammatic (FD) as well as in the effective potential (EP)
method -- as well as the \order{\alb\als}, 
\order{\alt\alb} and \order{\alb^2} contributions  -- evaluated in the EP
approach -- are known for vanishing external momenta.  
An evaluation of the momentum dependence at the two-loop level in a pure
\DRbar\ calculation was presented in \citere{mhiggs2lp2}.
A (nearly) full two-loop EP calculation,  
including even the leading three-loop corrections, has also been
published~\cite{mhiggsEP5}. However, the calculation presented in
\citere{mhiggsEP5} is not publicly available as a computer code 
for Higgs-mass calculations. 
Subsequently, another leading three-loop
calculation of \order{\alt\als^2}, depending on the various SUSY mass
hierarchies, has been performed~\cite{mhiggsFD3l},
resulting in the code {\tt H3m} which
adds the three-loop corrections to the 
{\tt FeynHiggs}~\cite{feynhiggs,mhiggslong,mhiggsAEC,mhcMSSMlong,Mh-logresum} 
result.
Recently, a combination of the full one-loop result, supplemented
with leading and subleading two-loop corrections evaluated in the
Feynman-diagrammatic/effective potential method and a resummation of the
leading and subleading logarithmic corrections from the scalar-top
sector has been published~\cite{Mh-logresum} in the latest version of
the code~\fh~\cite{feynhiggs,mhiggslong,mhiggsAEC,mhcMSSMlong,Mh-logresum}.

More recently the calculation of the momentum dependent two-loop
QCD corrections to $\Mh$ have been presented~\cite{Mh-p2-BH4}.
(From a technical point of view 
we have calculated the momentum dependent two-loop self-energy
diagrams numerically using 
the program {\tt SecDec}~\cite{Carter:2010hi,Borowka:2012yc,Borowka:2015mxa}.)
Subsequently, in \citere{Mh-p2-DDVS} this calculation was repeated
(differences of the two calculations are discussed in
\citere{Mh-p2-BH4-2}), where also a calculation of the two-loop
corrections of \order{\al\als} were presented.
The results of \citere{Mh-p2-BH4} are publicly available in the code \fh.

The remaining theoretical uncertainty in the
calculation of $\Mh$, from unknown higher-order corrections, had been
estimated  to be up to $3 \gev$, depending on the parameter region.
Recent improvements have potentially lead to a somewhat smaller estimate of 
up to $\sim 2 \gev$~\cite{Mh-logresum,ehowp} for not too large SUSY mass
scales. However, a careful re-analysis of this uncertainty for lower and
heavier SUSY mass scales is in order.
As the accuracy of the $\Mh$ prediction should at least match the one of the 
experimental result, further sub-dominant higher-order corrections 
have to be included in the Higgs-boson mass predictions~\cite{kuts}.


\section{Coupling strength analysis at the LHC and beyond}

Testing the coupling strengths of the discovered Higgs boson could yield
hints towards an extended Higgs sector, where the MSSM Higgs sector
makes clear predictions for possible deviations.
In order to test the compatibility of the predictions for the
SM Higgs boson with the (2012) experimental data, the LHC Higgs Cross
Section Working Group  proposed several benchmark scenarios for
``coupling scale
factors''~\cite{HiggsRecommendation,YR3} (see \citere{Englert:2014uua}
for a recent review on Higgs coupling extractions).
Effectively, the  predicted SM Higgs cross sections and partial
decay widths are dressed with scale factors $\kappa_i$ (and $\kappa_i=1$
corresponds to the SM).
Several assumptions are made for this $\kappa$-framework: there is only
one state at $\MHexp \gev$ responsible for the signal, the coupling
structure is the same as for the SM Higgs (i.e.\ it is a $\cp$-even
scalar), and the zero width approximation is assumed to be valid,
allowing for a clear separation and simple handling of production and
decay of the Higgs particle.  
The most relevant coupling strength modifiers are
$\kappa_t$, $\kappa_b$, $\kappa_\tau$, $\kappa_W$, $\kappa_Z$, 
$\kappa_\gamma$, $\kappa_g$, \ldots.

One limitation at the LHC (but not at the ILC) is the fact that the
total width cannot be determined experimentally without additional
theory assumptions. In the absence of a total width measurement only
ratios of $\kappa$'s can be determined from experimental data.
An assumption often made is $\kappa_{W,Z} \le 1$~\cite{Duhrssen:2004cv}.
A recent analysis from CMS using the Higgs decays to $ZZ$ far off-shell
yielded an upper limit on the total width about four times larger than the
SM width~\cite{CMS:2014ala}. However, here the assumption of the
equality of on-shell and off-shell couplings of the Higgs boson plays a
crucial role. It was pointed out that this equality is violated in
particular in the presence of new physics in the Higgs
sector~\cite{Englert:2014aca,Logan:2014ppa}.

\begin{figure}[htb!]
\begin{center}
  \includegraphics[width=0.45\textwidth]{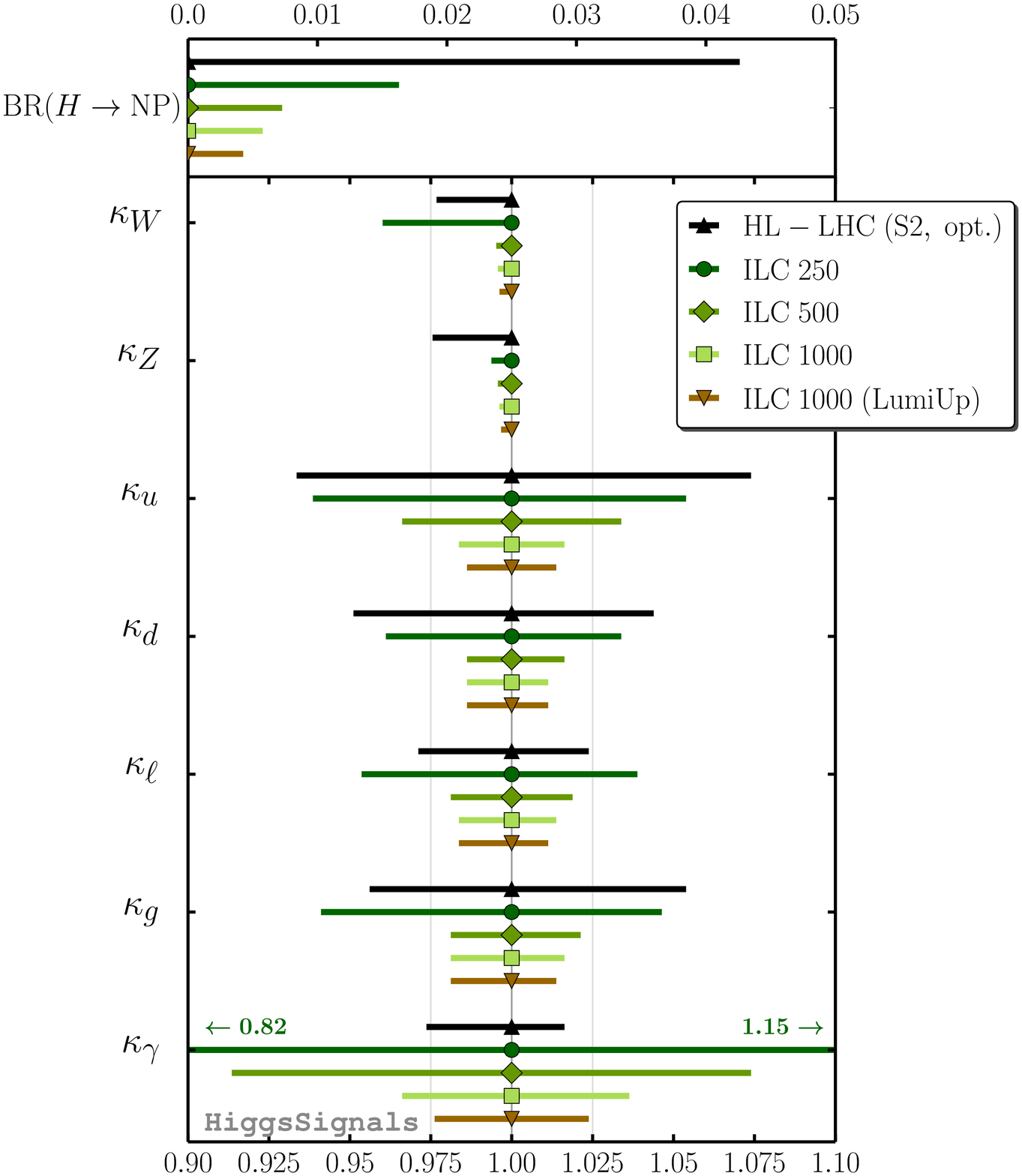}
  \includegraphics[width=0.45\textwidth]{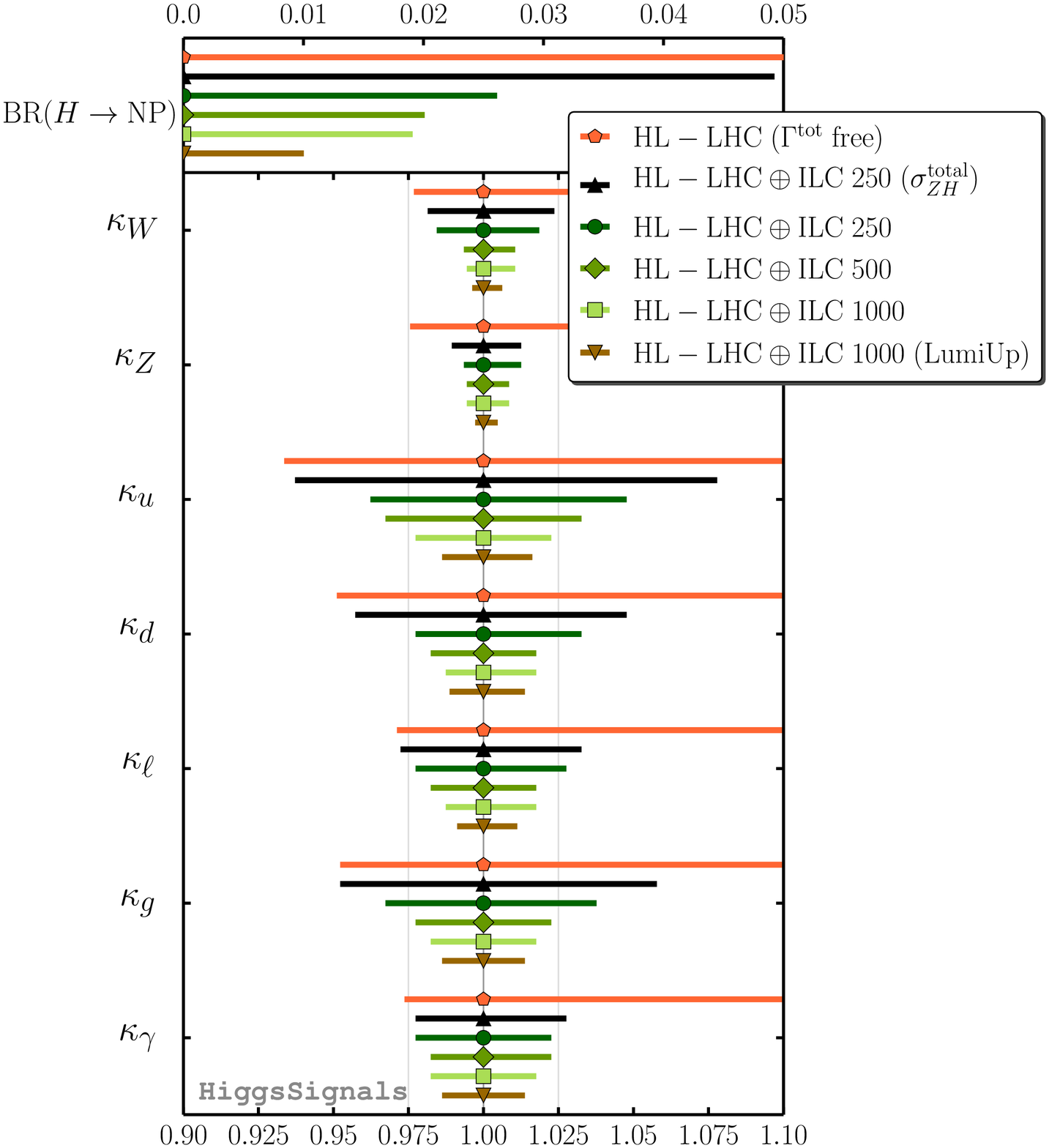}
\end{center}
\vspace{-1em}
\caption{Fit to the coupling scale factors with (left) and without the
  theory assumption of $\kappa_{W,Z} \le 1$~\cite{HiggsCouplings}.} 
  \label{fig:Hcoupl-ILC-LHC}
\end{figure}

In the left plot of \reffi{fig:Hcoupl-ILC-LHC} we compare the results
estimated for the HL-LHC (with $3 \iab$ and an assumed improvement of
50\% in the theoretical uncertainties) with the various stages of the ILC
under the theory
assumption $\kappa_{W,Z} \le 1$~\cite{HiggsCouplings}. This most
general fit includes $\kappa_{W,Z}$ for the gauge bosons,
$\kappa_{u,d,l}$ for up-type quarks, down-type quarks and charged
leptons, respectively, as well as $\kappa_\gamma$ and $\kappa_g$ for the
loop-induced couplings of the Higgs to photons and gluons. Also the
(possibly invisible) branching ratio of the Higgs boson to new physics
($\br(H \to {\rm NP})$) is included. One can observe that the HL-LHC
and the ILC\,250 yield comparable results. However, going to higher ILC
energies, yields substantially higher precisions in the fit for the
coupling scale factors. In the final stage of the ILC (ILC\,1000
LumiUp), precisions at the per-mille level in $\kappa_{W,Z}$ are
possible. The $1-2\%$ range is reached for all other $\kappa$'s.
The branching ratio to new physics can be restricted to the per-mille level.

Using ILC data the theory assumption $\kappa_{W,Z} \le 1$ can be
dropped, since the ``$Z$-recoil method'' (see \citere{Baer:2013cma} and
references therein)  
allows for a model independent determination of the $HZZ$ coupling. The
corresponding results are shown in the right plot
of \reffi{fig:Hcoupl-ILC-LHC}, where the HL-LHC results are combined
with the various stages of the ILC. The results from the HL-LHC alone 
continue to very large values of the $\kappa$'s, since the fit cannot be
done without theory assumptions.
Including the ILC measurements (where the first line corresponds to the
inclusion of {\em only} the $\si_{ZH}^{\rm total}$ measurement at the
ILC) yields a converging fit. In the final ILC stage $\kappa_{W,Z}$ are
determined 
to better than one per-cent, whereas the other coupling scale factors
are obtained in the $1-2\%$ range. The branching ratio to new physics is
restricted to be smaller than one per-cent. 
This opens up the possibility to observe MSSM induced deviations in the
Higgs boson couplings, provided that the overall Higgs mass scale,
$\MA$, is not too large.

\section{The search for additional Higgs bosons}

\begin{figure}[ht]
\begin{center}
\includegraphics[width=.48\textwidth,height=7.5cm]{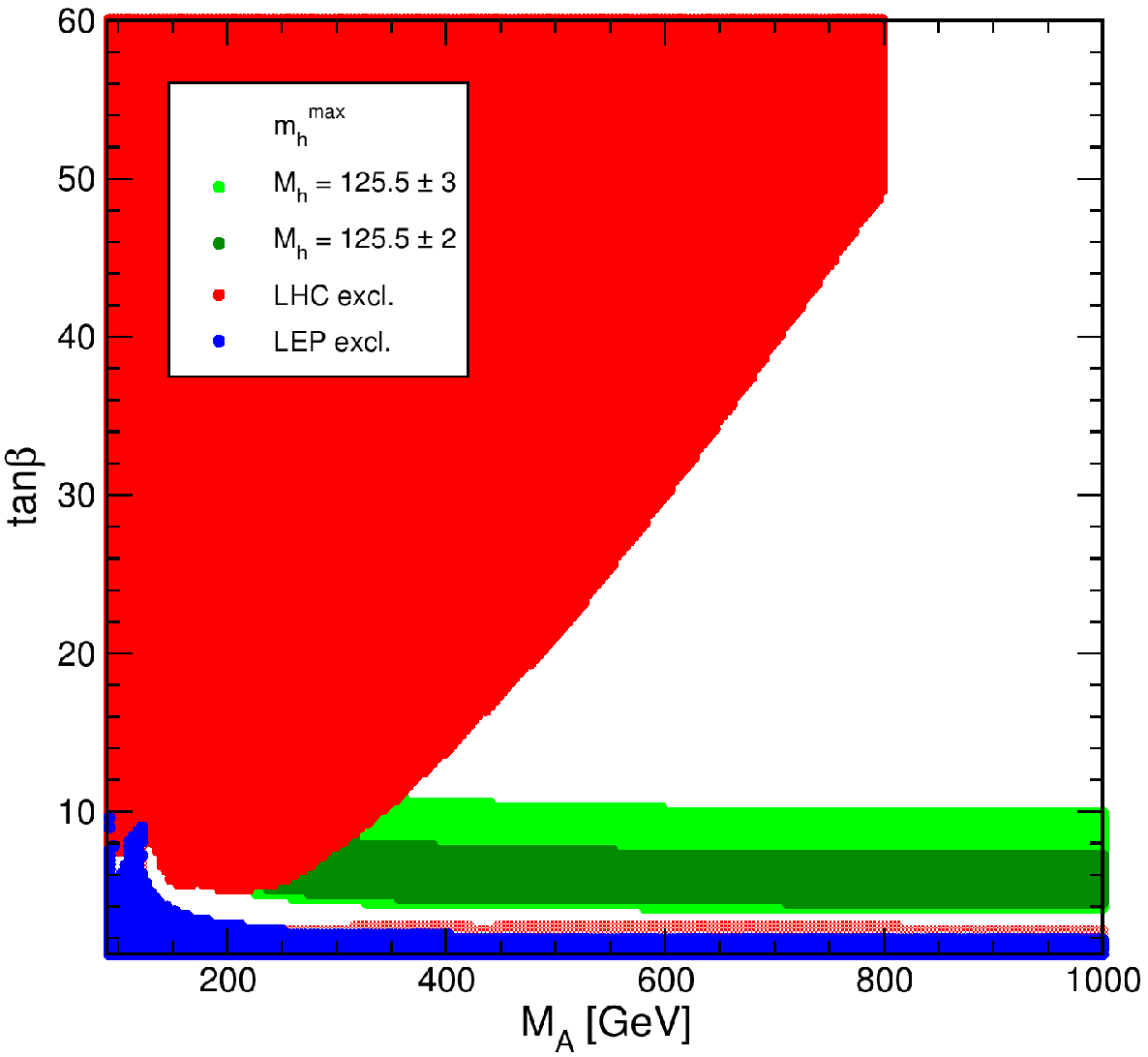}
\includegraphics[width=.48\textwidth,height=7.5cm]{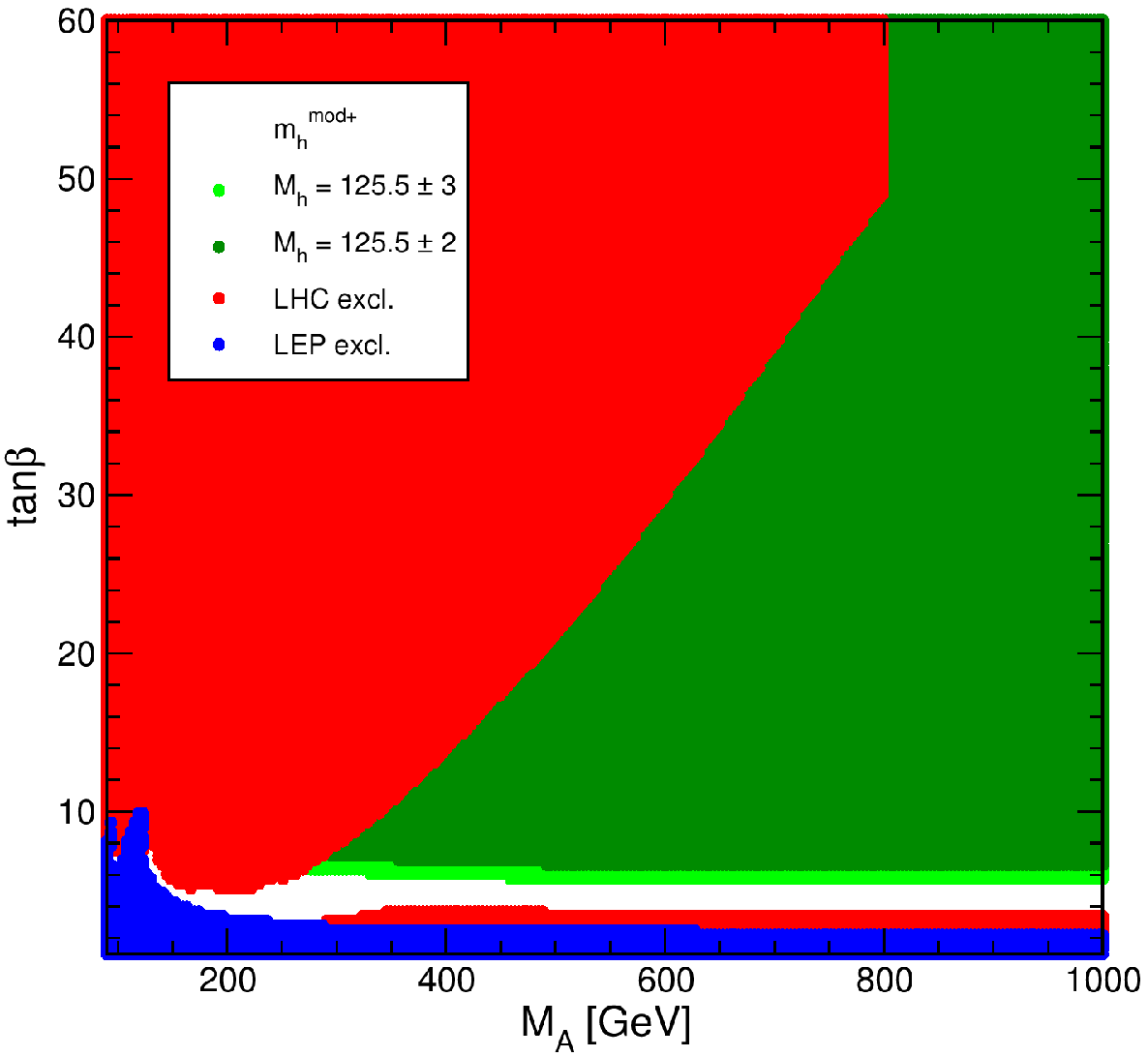}
\vspace{-1em}
\caption{%
$\MA$-$\tb$ plane in the $\mhmax$ scenario (left) and in the 
$\mh^{\rm mod+}$ scenario (right)~\cite{benchmark4}. 
The green shaded area
yields $\Mh \sim 125 \pm 3 \gev$, the red area at high $\tb$ is excluded
by LHC heavy MSSM Higgs boson searches, the blue area is excluded by LEP Higgs
searches, and the red strip at low $\tb$ is excluded by the LHC SM Higgs
searches.
}
\label{fig:benchmark4}
\end{center}
\end{figure}

Many investigations have been performed analyzing the agreement of the
MSSM with a Higgs boson at $\sim \MHexp \gev$. In a first step only the
mass information can be used to test the model, while in a second step
also the rate information of the various Higgs search channels can be
taken into account (see the previous section). 
Here we briefly review some results in two of the
new benchmark scenarios~\cite{benchmark4}, devised for the search for
heavy MSSM Higgs bosons. In the left plot of \reffi{fig:benchmark4} the
$\mhmax$ scenario is shown. The red area is excluded by LHC searches for
the heavy MSSM Higgs bosons, the blue area is excluded by LEP Higgs
searches, and the light shaded red area is excluded by LHC searches for
a SM-like Higgs boson. The bounds have been obtained with 
{\tt HiggsBounds}~\cite{higgsbounds} (where an extensive list of
original references can be found). The green area yields 
$\Mh = 125 \pm 3 \gev$, i.e.\ the region allowed by the experimental
data, taking into account the theoretical uncertainty in the $\Mh$
calculation as discussed above. 
Since the \mhmax\ scenario maximizes the light $\cp$-even Higgs boson
mass it is possible to extract lower (one parameter)
limits on $\MA$ and $\tb$ from the edges of the green band. 
By choosing the parameters entering via
radiative corrections such that those corrections yield a maximum upward
shift to $\Mh$, the lower bounds on $\MA$ and $\tb$ that can be
obtained are general in the sense that they (approximately) hold
for {\em any\/} values of the other parameters.
To address the (small) residual $\msusy$
dependence ($\msusy$ denotes the average scalar top mass scale)
of the lower bounds on $\MA$ and $\tb$, limits have been extracted
for the three different values $\msusy=\{0.5, 1, 2\}\tev$, see
\refta{tab:matblimits}~\cite{Mh125}. For comparison also
the previous limits derived from the LEP Higgs
searches~\cite{LEPHiggsMSSM} are shown, i.e.\  
before the incorporation of the Higgs discovery at the LHC. 
The bounds on $\MA$ translate directly into lower limits on $\MHp$,
which are also given in the table. More recent experimental Higgs
exclusion bounds shift these limits to even higher values, see the left
plot in \reffi{fig:benchmark4}. Consequently, the experimental result of 
$\Mh \sim \MHexp \pm 3 \gev$ requires $\MHp \gsim \mt$ with important
consequences for the charged Higgs boson phenomenology.

In the right plot of \reffi{fig:benchmark4} we show the $\mh^{\rm mod+}$
scenario that differs from the \mhmax\ scenario in the choice of
$\Xt$ (the off-diagonal entry in the scalar top mass matrix). 
While in the \mhmax\ scenario $\Xt/\msusy = +2$ had been chosen to
maximize $\Mh$, in the $\mh^{\rm mod+}$ scenario $\Xt/\msusy = +1.5$ is
used to yield a ``good'' $\Mh$ value over the nearly the entire
$\MA$-$\tb$ plane, which is visible as the extended green region. 

\begin{table}[ht]
\centering
\begin{tabular}{c|ccc|ccc}
\hline
& \multicolumn{3}{c|}{Limits without $\Mh\sim125\gev$} & \multicolumn{3}{c}{Limits with $\Mh\sim125\gev$}\\
$\msusy$ (GeV) & $\tb$ & $\MA$ (GeV) & $\MHp$ (GeV) & $\tb$& $\MA$ (GeV) & $\MHp$ (GeV)  \\
\hline
500 & $2.7$ & $95$ & $123$ & $4.5$ & $140$ & $161$\\
1000 & $2.2$ & $95$ & $123$ & $3.2$ & $133$ & $155$ \\
2000& $2.0$ & $95$ & $123$ & $2.9$ & $130$ & $152$\\
\hline
\end{tabular}
\caption{Lower limits on the MSSM Higgs sector tree-level parameters
  $\MA$ ($\MHp$) and $\tb$ obtained with and without the assumed Higgs
  signal of $\Mh \sim125.5 \gev$. The mass limits have been rounded to
  $1 \gev$~\cite{Mh125}.}  
\label{tab:matblimits}
\end{table}


\section{Precision predictions for the decay of Higgs bosons to SUSY
  particles} 

\begin{figure}[ht]
\begin{center}
\includegraphics[width=.48\textwidth,height=7.0cm]{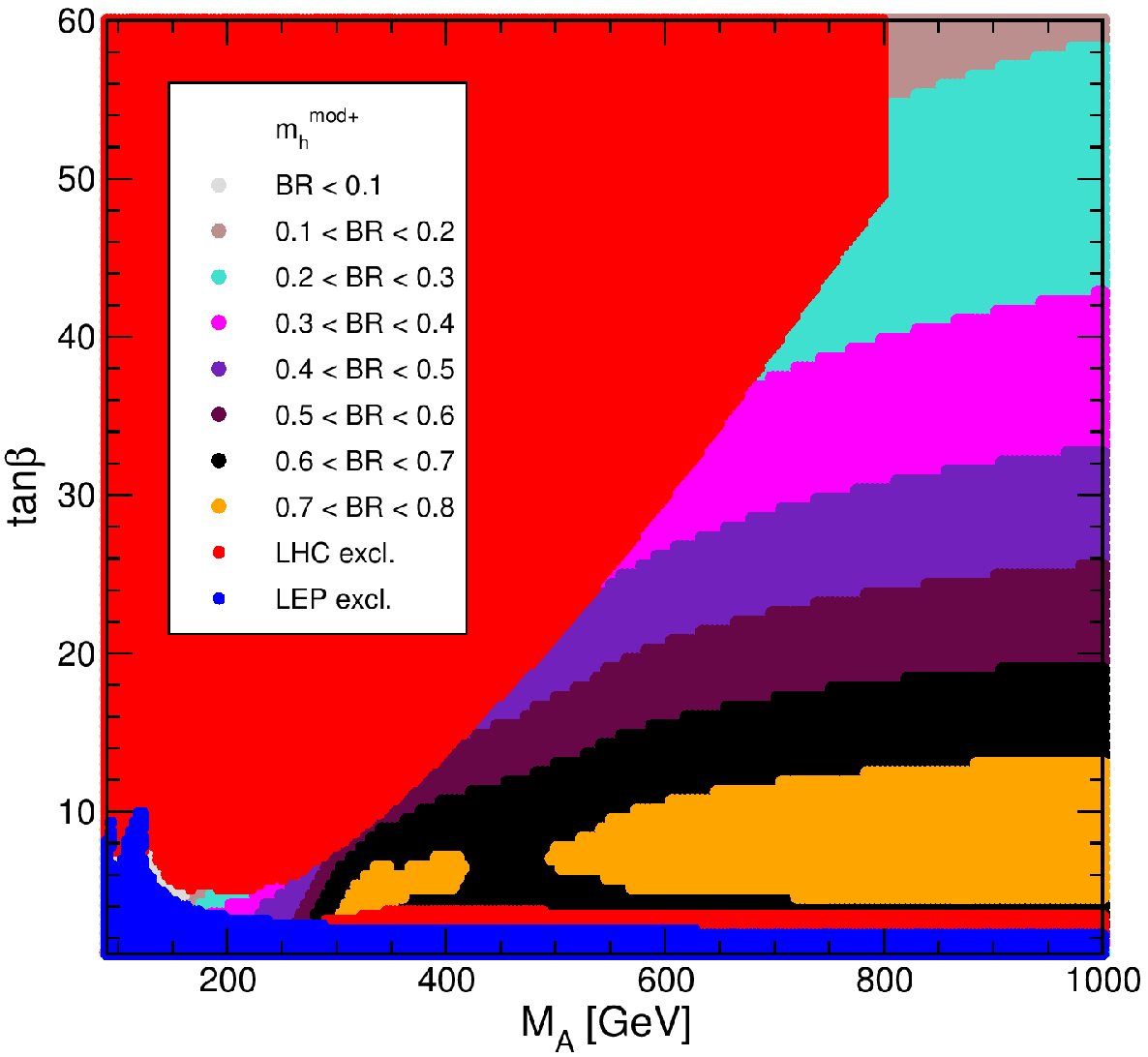}
\includegraphics[width=.48\textwidth,height=7.0cm]{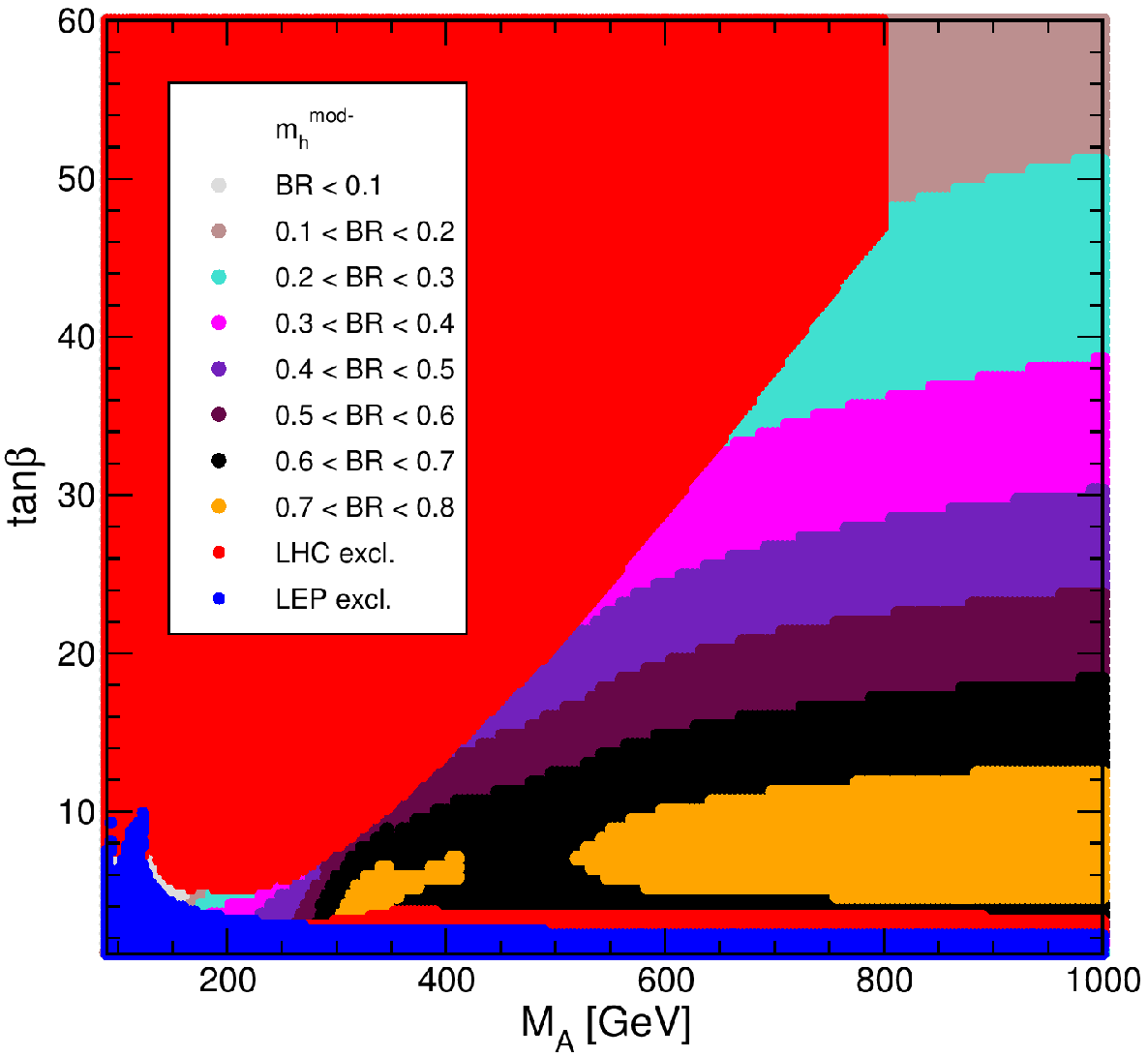}
\vspace{-1em}
\caption{%
$\MA$-$\tb$ plane in the $\mh^{\rm mod+}$ scenario (left)
and the $\mh^{\rm mod-}$ scenario (right)~\cite{benchmark4}. 
he exclusion regions are shown as in \reffi{fig:benchmark4},
while the color coding in the allowed region indicates the average total 
branching ratio of $H$ and $A$ into 
charginos and neutralinos.
}
\label{fig:benchmark4-cn}
\end{center}
\end{figure}

Depending on the scale of Higgs and SUSY masses the main decay channels
of the additional Higgs bosons could go to SUSY particles, which 
is demonstrated in \reffi{fig:benchmark4-cn}~\cite{benchmark4}. 
The branching ratios for the decay of $H$ and $A$ into 
charginos and neutralinos may become large at
small or moderate values of $\tb$. 
In \reffi{fig:benchmark4-cn} we show the \mhmodp\ (left) and \mhmodm\
(right) scenarios~\cite{benchmark4}, where the masses of the charginos
and neutralinos are \order{200 \gev}.
The excluded regions from the
Higgs searches at LEP and the LHC are as before. 
The color coding for the allowed region of the parameter space
indicates the average value of the branching ratios for the decay of $H$
and $A$ into charginos and neutralinos (summed over all contributing
final states).
One can see from the plots that as a consequence of the
relatively low values of the chargino/neutralino masses in these
benchmark scenarios the 
decays of $H$ and $A$ into charginos and neutralinos are kinematically
open essentially in the whole allowed parameter space of the scenario,
with the exception of a small region with rather small $\MA$. The
branching ratios for the decays of $H$ and $A$ into charginos and
neutralinos reach values in excess of 70\% for small and moderate values of
$\tb$. Including these channels into the searches for heavy MSSM Higgs
bosons could potentially allow to discover new Higgs bosons and SUSY
particles at the same time.

\begin{figure}[ht!]
\begin{center}
\begin{tabular}{c}
\includegraphics[width=0.49\textwidth,height=7.0cm]{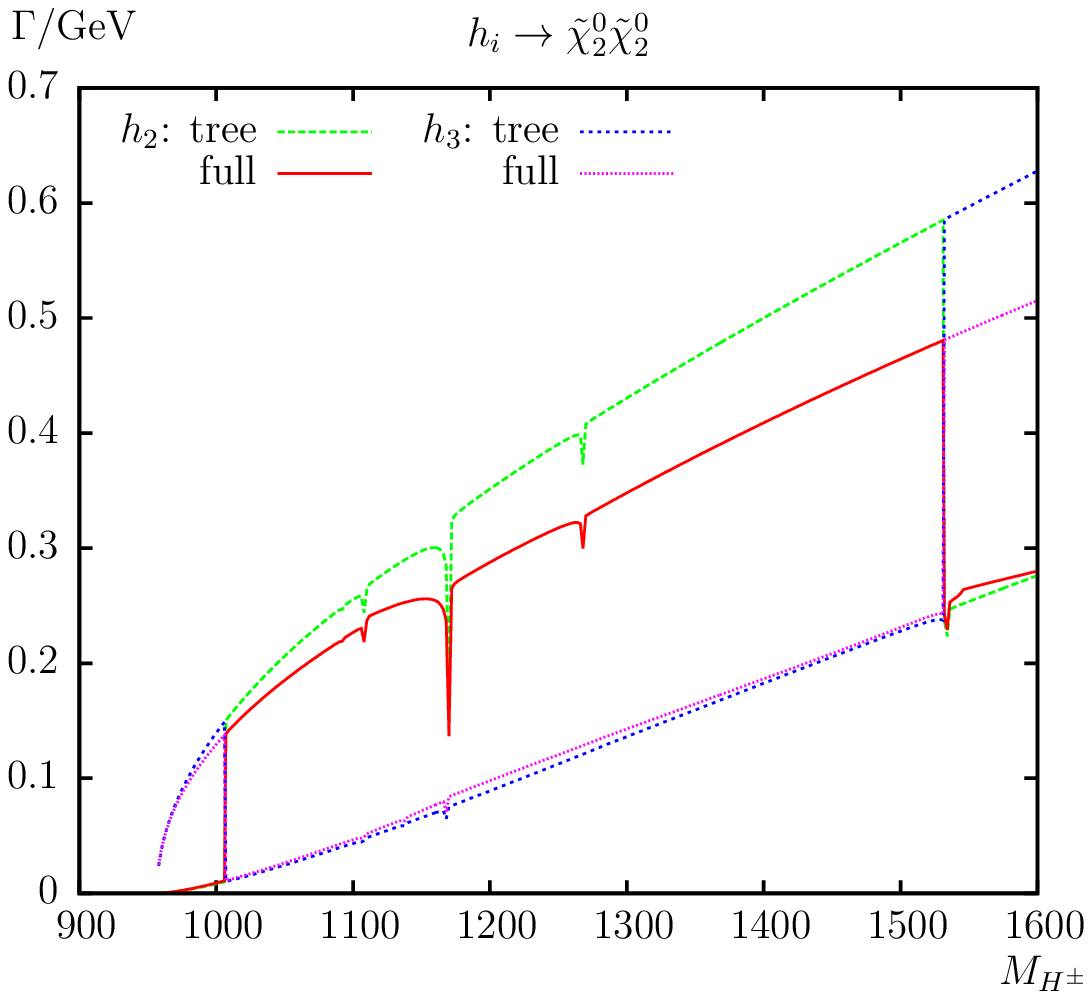}
\hspace{-4mm}
\includegraphics[width=0.49\textwidth,height=7.0cm]{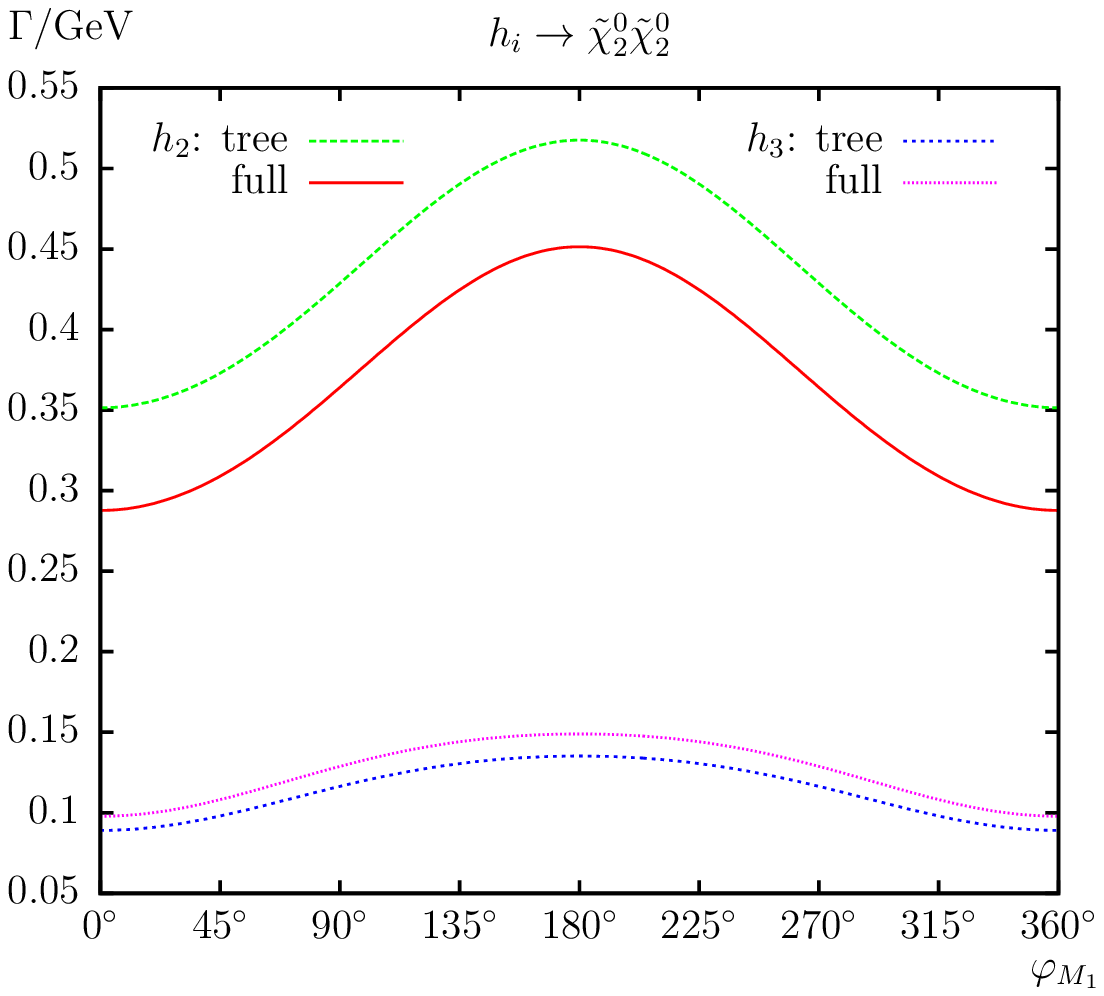}
\end{tabular}
\vspace{-1em}
\caption{
  $\Ga(h_i \to \neu2\neu2)$. 
  Tree-level and full one-loop corrected partial widths for 
  $h_i \to \neu2\neu2$ ($i = 2,3$) are shown~\cite{HiggsDecayIno}.
  The left plot shows the partial decay width with $\MHp$ varied. 
  The right plot shows the complex phase $\phiMe$ varied (see text).
}
\label{fig:hneu2neu2}
\end{center}
\vspace{-1em}
\end{figure}

Recently,
full one-loop calculations for the decays of Higgs bosons to scalar
fermions~\cite{HiggsDecaySferm} and into
charginos/neutralinos~\cite{HiggsDecayIno} in the MSSM with complex
parameters (cMSSM) have become available.
In \reffi{fig:hneu2neu2} we show the results for the decay 
$h_i \to \neu2\neu2$ ($i = 2,3$), where details and parameter settings
can be found in \citere{HiggsDecayIno}. $h_2$ and $h_3$ are the two
neutral heavy Higgs bosons in the cMSSM,
corresponding to $H$ and $A$ in the real case. In the left plot of
\reffi{fig:hneu2neu2} the decay widths at the tree- and at the full
one-loop level are shown as a function of
$\MHp$ (at $\MHp \sim 1000 \gev$ and $\MHp \sim 1520 \gev$ a mass
crossing of $h_2$ and $h_3$ takes place, see \citere{HiggsDecayIno} for
details). It can be seen that the (in this case purely electroweak)
one-loop correction can change the 
decay width by up to $20\%$. In the right plot of \reffi{fig:hneu2neu2}
the decay widths are shown as a function of $\phiMe$, the phase of the
$U(1)$ gaugino soft SUSY-breaking parameter. Changing the phase can lead
to effects of up to $50\%$, while the one-loop corrections again can be
as large as $20\%$. These examples show that complex parameters and the full
one-loop corrections should be taken into account for the interpretation of the 
searches for charginos/neutralinos as well as for any future precision 
analyses of those decays.

\begin{acknowledgments}

I thank the organizers of the HPNP2015 for the invitation, the very 
stimulating workshop and the excellent Japanese food. 
I felt very well taken care of. Thanks!

\end{acknowledgments}


\bigskip 


\begin{thebibliography}{99} 

\bibitem{ATLASdiscovery} G.~Aad et al. [ATLAS Collaboration],
  {\em Phys.\ Lett.} {\bf B 716} (2012) 1
  [arXiv:1207.7214 [hep-ex]].

\bibitem{CMSdiscovery} S.~Chatrchyan et al. [CMS Collaboration],
  {\em Phys.\ Lett.} {\bf B 716} (2012) 30
  [arXiv:1207.7235 [hep-ex]].

\bibitem{Moriond15} 
  M.~D\"uhrssen, talk given at `Rencontres de Moriond EW 2014'', see:\\
{\tt https://indico.in2p3.fr/event/10819/session/3/contribution/102/material/slides/1.pdf}~.

\bibitem{mssm} H.~Nilles, 
               {\em Phys.\ Rept.} {\bf 110} (1984) 1; 
               H.~Haber and G.~Kane, 
               {\em Phys.\ Rept.} {\bf 117} (1985) 75; 
               R.~Barbieri, 
               {\em Riv.\ Nuovo Cim.} {\bf 11} (1988) 1. 

\bibitem{MHreviews} A.~Djouadi,
                    {\em Phys.\ Rept.} {\bf 459} (2008) 1
                    [arXiv:hep-ph/0503173];
                    S.~Heinemeyer,
                    {\em Int.\ J.\ Mod.\ Phys.} {\bf A 21} (2006) 2659
                    [arXiv:hep-ph/0407244].

\bibitem{PomssmRep} S.~Heinemeyer, W.~Hollik and G.~Weiglein,
                    {\em Phys.\ Rept.} {\bf 425} (2006) 265
                    [arXiv:hep-ph/0412214].

\bibitem{mhiggsAEC} G.~Degrassi, S.~Heinemeyer, W.~Hollik,
                    P.~Slavich and G.~Weiglein, 
                    {\em Eur. Phys. J.} {\bf C 28} (2003) 133
                    [arXiv:hep-ph/0212020].

\bibitem{ATLASCMS-MH} G.~Aad et al. [ATLAS and CMS Collaborations],
  arXiv:1503.07589 [hep-ex].

\bibitem{ERZ} J.~Ellis, G.~Ridolfi and F.~Zwirner,
              {\em Phys.\ Lett.} {\bf B 257} (1991) 83;
              Y.~Okada, M.~Yamaguchi and T.~Yanagida,
              {\em Prog.\ Theor.\ Phys. } {\bf 85} (1991) 1;
              H.~Haber and R.~Hempfling,
              {\em Phys.\ Rev.\ Lett.}  {\bf 66} (1991) 1815.

\bibitem{mhiggsf1lABC} A.~Brignole,
                     {\em Phys. Lett.}\ {\bf B 281} (1992) 284;
                     P.~Chankowski, S.~Pokorski and J.~Rosiek,
                     {\em Phys. Lett.} {\bf B 286} (1992) 307;
                     {\em Nucl. Phys.} {\bf B 423} (1994) 437
                     [arXiv:hep-ph/9303309];
                     A.~Dabelstein,
                     {\em Nucl. Phys.} {\bf B 456} (1995) 25
                     [arXiv:hep-ph/9503443];
                     {\em Z. Phys.} {\bf C 67} (1995) 495
                     [arXiv:hep-ph/9409375].

\bibitem{mhiggsletter} S.~Heinemeyer, W.~Hollik and G.~Weiglein, 
                       {\em Phys. Rev.} {\bf D 58} (1998) 091701
                       [arXiv:hep-ph/9803277]; 
                       {\em Phys. Lett.} {\bf B 440} (1998) 296
                       [arXiv:hep-ph/9807423].

\bibitem{mhiggslong} S.~Heinemeyer, W.~Hollik and G.~Weiglein,
                     {\em Eur. Phys. J.} {\bf C 9} (1999) 343
                     [arXiv:hep-ph/9812472].

\bibitem{mhiggslle} S.~Heinemeyer, W.~Hollik and G.~Weiglein,
                    {\em Phys. Lett.} {\bf B 455} (1999) 179
                    [arXiv:hep-ph/9903404].

\bibitem{mhiggsFD2} S.~Heinemeyer, W.~Hollik, H.~Rzehak and G.~Weiglein,
                    {\em Eur. Phys. J.} {\bf C 39} (2005) 465
                    [arXiv:hep-ph/0411114].

\bibitem{bse} M.~Carena et al., 
              {\em Nucl. Phys.} {\bf B 580} (2000) 29
              [arXiv:hep-ph/0001002].

\bibitem{mhiggsEP0}  R.~Zhang, 
                     {\em Phys.\ Lett. } {\bf B 447} (1999) 89
                     [arXiv:hep-ph/9808299];
                     J.~Espinosa and R.~Zhang, 
                     {\em JHEP} {\bf 0003} (2000) 026
                     [arXiv:hep-ph/9912236].

\bibitem{mhiggsEP1} G.~Degrassi, P.~Slavich and F.~Zwirner,
                    {\em Nucl. Phys.} {\bf B 611} (2001) 403
                    [arXiv:hep-ph/0105096].

\bibitem{mhiggsEP1b} R.~Hempfling and A.~Hoang, 
                     {\em Phys. Lett.} {\bf B 331} (1994) 99
                     [arXiv:hep-ph/9401219].

\bibitem{mhiggsEP2} A.~Brignole, G.~Degrassi, P.~Slavich and F.~Zwirner,
                    {\em Nucl. Phys.} {\bf B 631} (2002) 195
                    [arXiv:hep-ph/0112177].

\bibitem{mhiggsEP3} J.~Espinosa and R.~Zhang,
                    {\em Nucl. Phys.} {\bf B 586} (2000) 3
                    [arXiv:hep-ph/0003246]. 

\bibitem{mhiggsEP3b} J.~Espinosa and I.~Navarro,
                     {\em Nucl.\ Phys.} {\bf B 615} (2001) 82
                     [arXiv:hep-ph/0104047].

\bibitem{mhiggsEP4} A.~Brignole, G.~Degrassi, P.~Slavich and F.~Zwirner,
                    {\em Nucl. Phys.} {\bf B 643} (2002) 79
                    [arXiv:hep-ph/0206101].

\bibitem{mhiggsEP4b} G.~Degrassi, A.~Dedes and P.~Slavich,
                    {\em Nucl. Phys.} {\bf B 672} (2003) 144
                    [arXiv:hep-ph/0305127].

\bibitem{mhiggsRG1} M.~Carena, J.~Espinosa, M.~Quir\'os and C.~Wagner, 
                    {\em Phys. Lett.} {\bf B 355} (1995) 209
                    [arXiv:hep-ph/9504316];\\
                    M.~Carena, M.~Quir\'os and C.~Wagner, 
                    {\em Nucl. Phys.} {\bf B 461} (1996) 407
                    [arXiv:hep-ph/9508343].

\bibitem{mhiggsRG1a} J.~Casas, J.~Espinosa, M.~Quir\'os and A.~Riotto,
                     {\em Nucl. Phys.} {\bf B 436} (1995) 3,
                     [Erratum-ibid.\ {\bf B 439} (1995) 466]
                     [arXiv:hep-ph/9407389].

\bibitem{mhiggsFDalt2} W.~Hollik and S.~Pa\ss ehr,
  {\em JHEP} {\bf 1410} (2014) 171
  [arXiv:1409.1687 [hep-ph]].

\bibitem{mhiggs2lp2} S.~Martin,
                    {\em Phys. Rev.} {\bf D 71} (2005) 016012
                    [arXiv:hep-ph/0405022].

\bibitem{mhiggsEP5} S.~Martin, 
                    {\em Phys. Rev.} {\bf D 65} (2002) 116003
                    [arXiv:hep-ph/0111209];
                    {\bf D 66} (2002) 096001
                    [arXiv:hep-ph/0206136];
                    {\bf D 67} (2003) 095012
                    [arXiv:hep-ph/0211366];
                    {\bf D 68} (2003) 075002
                    [arXiv:hep-ph/0307101]; 
                    {\bf D 70} (2004) 016005
                    [arXiv:hep-ph/0312092];
                    {\bf D 71} (2005) 116004
                    [arXiv:hep-ph/0502168];
                    {\bf D 75} (2007) 055005
                    [arXiv:hep-ph/0701051];
                    S.~Martin and D.~Robertson,
                    {\em Comput.\ Phys.\ Commun.} {\bf 174} (2006) 133
                    [arXiv:hep-ph/0501132].

\bibitem{mhiggsFD3l} R.~Harlander, P.~Kant, L.~Mihaila and M.~Steinhauser,
                     {\em Phys.\ Rev.\ Lett.} {\bf 100} (2008) 191602
                     [{\em Phys.\ Rev.\ Lett.} {\bf 101} (2008) 039901]
                     [arXiv:0803.0672 [hep-ph]];
                     {\em JHEP} {\bf 1008} (2010) 104
                     [arXiv:1005.5709 [hep-ph]].

\bibitem{feynhiggs} S.~Heinemeyer, W.~Hollik and G.~Weiglein,
                    {\em Comput. Phys. Commun.} {\bf 124} (2000) 76
                    [arXiv:hep-ph/9812320];
                    T.~Hahn et al., 
                    {\em Comput.\ Phys.\ Commun.} {\bf 180} (2009) 1426; 
                    see: {\tt www.feynhiggs.de} .

\bibitem{mhcMSSMlong} M.~Frank et al, 
                      {\em JHEP} {\bf 0702} (2007) 047
                      [arXiv:hep-ph/0611326].

\bibitem{Mh-logresum} T.~Hahn, S.~Heinemeyer, W.~Hollik, H.~Rzehak and
                      G.~Weiglein, 
                      {\em Phys. Rev. Lett.} {\bf 112} (2014) 14, 141801
                      [arXiv:1312.4937 [hep-ph]].

\bibitem{Mh-p2-BH4} S.~Borowka, T.~Hahn, S.~Heinemeyer, G.~Heinrich and
  W.~Hollik, 
  {\em Eur.\ Phys.\ J.} {\bf C 74} (2014) 8,  2994
  [arXiv:1404.7074 [hep-ph]].

\bibitem{Carter:2010hi}
  J.~Carter and G.~Heinrich,
  {\em Comput.\ Phys.\ Commun.} {\bf 182} (2011) 1566
  [arXiv:1011.5493 [hep-ph]].

\bibitem{Borowka:2012yc}
  S.~Borowka, J.~Carter and G.~Heinrich,
  {\em Comput.\ Phys.\ Commun.} {\bf 184} (2013) 396
  [arXiv:1204.4152 [hep-ph]].

\bibitem{Borowka:2015mxa}
  S.~Borowka, G.~Heinrich, S.~Jones, M.~Kerner, J.~Schlenk and T.~Zirke,
  arXiv:1502.06595 [hep-ph].

\bibitem{Mh-p2-DDVS} G.~Degrassi, S.~Di Vita and P.~Slavich,
  {\em Eur.\ Phys.\ J.} {\bf C 75} (2015) 2,  61
  [arXiv:1410.3432 [hep-ph]].

\bibitem{Mh-p2-BH4-2} S.~Borowka, T.~Hahn, S.~Heinemeyer, G.~Heinrich
                      and W.~Hollik,
                      MPP--2015--97, ZU-TH 11/15, 
                      {\em in preparation}.

\bibitem{ehowp} O.~Buchmueller et al., 
                {\em Eur.\ Phys.\ J.} {\bf C 74} (2014) 3,  2809
                [arXiv:1312.5233 [hep-ph]].

\bibitem{kuts} See: {\tt https://sites.google.com/site/kutsmh}~.

\bibitem{HiggsRecommendation} LHC Higgs Cross Section Working Group,
  A.~David et al., 
  arXiv:1209.0040 [hep-ph].

\bibitem{YR3} S.~Heinemeyer et al.
              [LHC Higgs Cross Section Working Group],
              arXiv:1307.1347 [hep-ph].

\bibitem{Englert:2014uua}
  C.~Englert et al., 
  {\em J.\ Phys.} {\bf G 41} (2014) 113001
  [arXiv:1403.7191 [hep-ph]].

\bibitem{Duhrssen:2004cv}
  M.~D\"uhrssen et al., 
  {\em Phys.\ Rev.} {\bf D 70} (2004) 113009
  [arXiv:hep-ph/0406323].

\bibitem{CMS:2014ala}
  CMS Collaboration,
  CMS-PAS-HIG-14-002.

\bibitem{Englert:2014aca}
  C.~Englert and M.~Spannowsky,
  {\em Phys.\ Rev.} {\bf D 90} (2014) 5,  053003
  [arXiv:1405.0285 [hep-ph]].

\bibitem{Logan:2014ppa}
  H.~Logan,
  arXiv:1412.7577 [hep-ph].

\bibitem{HiggsCouplings} P.~Bechtle 
S.~Heinemeyer, O.~St{\aa}l, T.~Stefaniak and G.~Weiglein,
                        {\em JHEP} {\bf 1411} (2014) 039
                        [arXiv:1403.1582 [hep-ph]];
  {\em Eur.\ Phys.\ J.} {\bf C 74} (2014) 2,  2711
  [arXiv:1305.1933 [hep-ph]].

\bibitem{Baer:2013cma}
  H.~Baer et al.,
  arXiv:1306.6352 [hep-ph].

\bibitem{benchmark4}
  M.\,Carena,\,S.\,Heinemeyer,\,O.\,St{\aa}l,\,C.\,Wagner,\,G.\,Weiglein,
  {\em Eur.\  Phys.\  J.} {\bf C\,73}\,(2013)\,2552
  [arXiv:1302.7033 [hep-ph]].

\bibitem{higgsbounds} P.~Bechtle et al., 
  {\em Comput.\ Phys.\ Commun.} {\bf 181} (2010) 138
  [arXiv:0811.4169 [hep-ph]];
  {\bf 182} (2011) 2605
  [arXiv:1102.1898 [hep-ph]];
  P.~Bechtle et al.,
  {\em Eur.\ Phys.\ J.} {\bf C 74} (2014) 2693
  [arXiv:1311.0055 [hep-ph]].

\bibitem{Mh125} S.~Heinemeyer, O.~St{\aa}l and G.~Weiglein, 
                {\em Phys.\ Lett.} {\bf B 710} (2012) 201
                [arXiv:1112.3026 [hep-ph]]; 

\bibitem{LEPHiggsMSSM} LEP Higgs working group,
                       {\em Eur.\ Phys.\ J.} {\bf C 47} (2006) 547
                       [arXiv:hep-ex/0602042].

\bibitem{HiggsDecaySferm} S.~Heinemeyer and C.~Schappacher,
  to appear in {\em Eur. Phys. J.} {\bf C}, 
  arXiv:1410.2787 [hep-ph]. 

\bibitem{HiggsDecayIno} S.~Heinemeyer and C.~Schappacher,
  arXiv:1503.02996 [hep-ph].



\end{thebibliography}
\end{document}